\documentclass[11pt]{iopart}

\usepackage{epsfig}
\def\laq{\raise 0.4ex\hbox{$<$}\kern -0.8em\lower 0.62ex\hbox{$\sim$}}
\def\gaq{\raise 0.4ex\hbox{$>$}\kern -0.7em\lower 0.62ex\hbox{$\sim$}}
\newcommand{\beq}{\begin{equation}} 
\newcommand{\eeq}{\end{equation}}
\newcommand{\bea}{\begin{eqnarray}} 
\newcommand{\eea}{\end{eqnarray}}

\begin{document}

\title[Laser-interferometer gravitational-wave 
optical-spring detectors]{Laser-interferometer gravitational-wave 
optical-spring detectors}

\author{Alessandra Buonanno\dag\ddag~ and Yanbei Chen\ddag}

\address{\dag Institut d'Astrophysique de Paris (GReCO, FRE 2435 du CNRS), 
98$^{\rm bis}$ Boulevard Arago, 75014 Paris, France} 
\address{\ddag Theoretical Astrophysics and 
Relativity Group, California Institute of Technology, 
Pasadena, 91125, CA, USA}

\begin{abstract}
Using a quantum mechanical approach, we show that in a 
gravitational-wave interferometer composed of arm cavities 
and a signal recycling cavity, e.g., the LIGO-II configuration,
the radiation-pressure force acting on the mirrors 
not only disturbs the motion of the free masses randomly due 
to quantum fluctuations, 
but also and more fundamentally, makes them respond to 
forces as though they were connected to an (optical) spring with 
a specific rigidity. This oscillatory response gives rise 
to a much richer dynamics than previously known, which 
enhances the possibilities for reshaping the LIGO-II's 
noise curves. However, the optical-mechanical system is 
dynamically unstable and an appropriate control system 
must be introduced to quench the instability.
\end{abstract}

\ead{buonanno@tapir.caltech.edu, yanbei@tapir.caltech.edu}

\vspace{8cm}

\noindent
{\small Contributed talk given at the $4^{th}$ Edoardo Amaldi 
Conference on Gravitational Waves, Perth, Australia, 8-13 July 
2001, to appear in Special Issue Article of Classical and Quantum Gravity}
\maketitle

\section{Introduction}
\label{sec1}

A network of broadband ground-based laser interferometers,
aimed to detect gravitational waves (GWs) in the frequency band
$10-10^4\,$Hz, will begin operations next year. 
This network is composed of GEO, the Laser Interferometer 
Gravitational-wave Observatory (LIGO), TAMA and VIRGO 
(whose operation will begin in 2004)~\cite{Inter}.
The LIGO Scientific Collaboration (LSC) \cite{GSSW99} is 
currently planning an upgrade of LIGO starting from 2007. 
Besides the improvement of the seismic isolation and 
suspension systems, and the increase (decrease) of light power 
(shot noise) circulating in the arm cavities, the LIGO community  
has planned to introduce an extra mirror, called a 
signal-recycling (SR) mirror~\cite{SRt}, 
at the dark-port output (see Fig.~\ref{Fig1}). 
The optical system composed of the SR cavity and the arm 
cavities forms a composite resonant cavity, whose 
eigenfrequencies and quality factors can be controlled by
the position and reflectivity of the SR mirror.
These eigenfrequencies (resonances) can be exploited 
to reshape the noise curves, enabling the interferometer to work
either in broadband or in narrowband
configurations, and improving in this way
the observation of specific GW astrophysical
sources~\cite{KT}. 

The initial theoretical analyses~\cite{SRt} and 
experiments~\cite{SRe} of SR interferometers 
refer to configurations with low laser power, 
for which the radiation pressure on the
arm-cavity mirrors is negligible and the quantum-noise spectra
are dominated by shot noise.
However, when the laser power is increased, the shot noise decreases 
while the effect of radiation-pressure fluctuation increases. 
LIGO-II has been planned to work at a laser power for which the
two effects are comparable in the observation band $10$--$200$\,Hz \cite{GSSW99}. 
Therefore, to correctly describe the quantum optical noise in LIGO-II, 
the results so far obtained in the literature had to 
be complemented by a thorough investigation of the influence of the
radiation-pressure force on the mirror motion.
Using a quantum-mechanical approach~\cite{BK92,KLMTV} we have recently investigated
~\cite{BC1,BC2,BC3} this issue. 
Henceforth, we shall summarize the main results of our analysis.

\section{Radiation-pressure forces in conventional versus signal-recycling interferometers }
\label{sec2}

\begin{figure}
\begin{center}
\begin{tabular}{cc}
\hspace{-0.4cm}
\epsfig{file=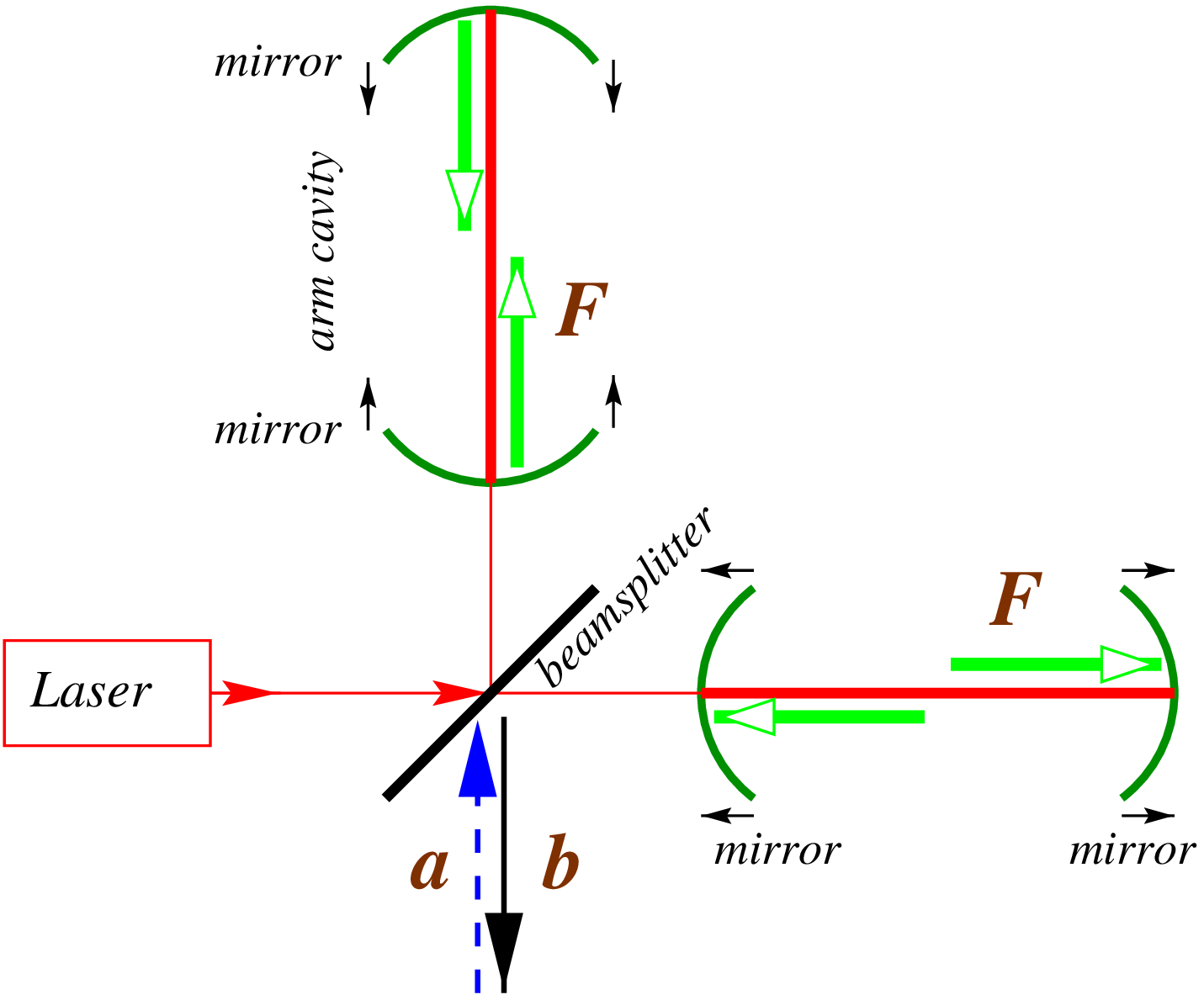,width=0.45\textwidth,angle=-90} 
& \epsfig{file=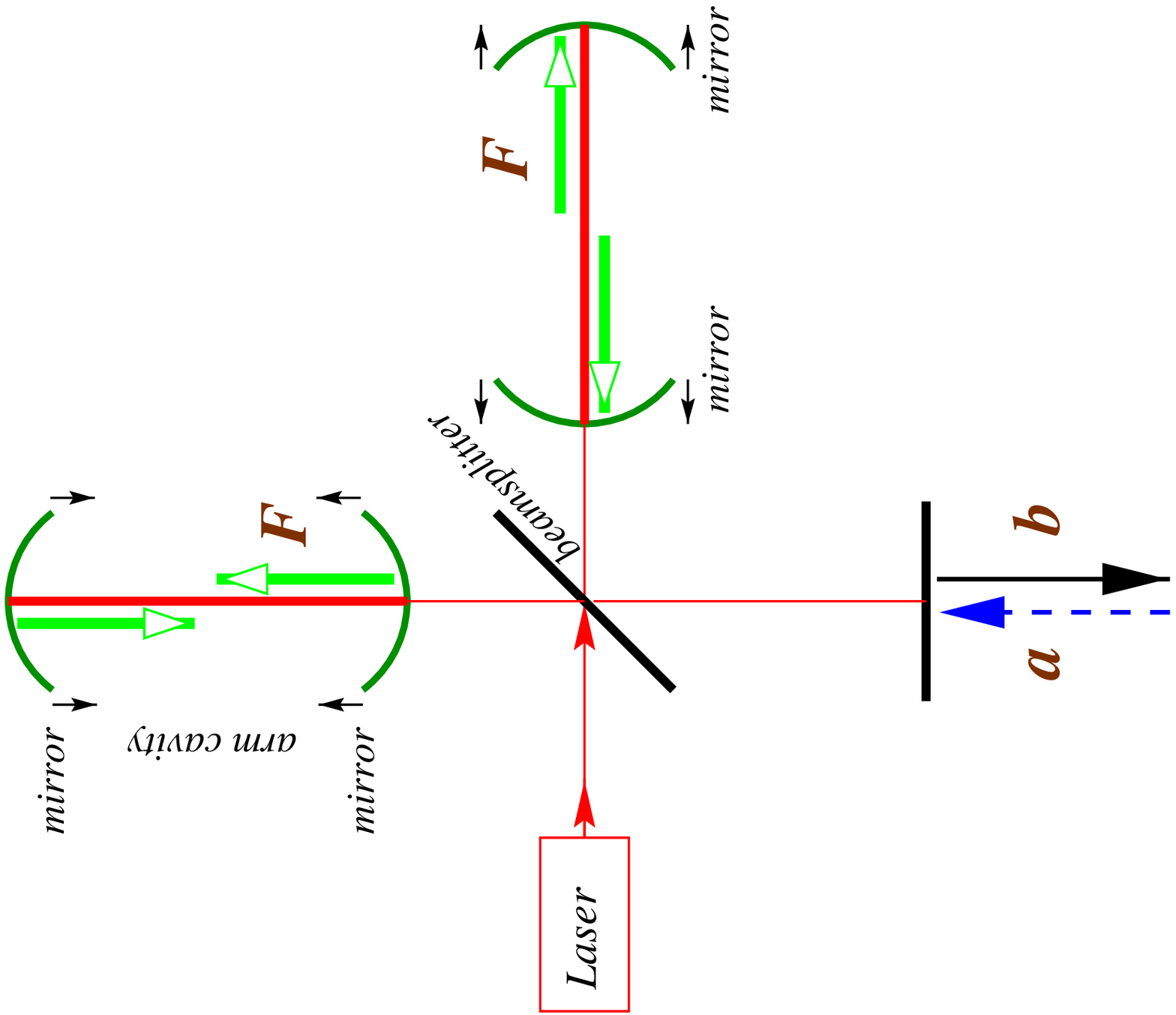,width=0.45\textwidth,angle=-90} 
\end{tabular}
\caption{Schematic views of a conventional interferometer 
(on the left panel) and of a signal-recycling interferometer 
(on the right panel).}
\label{Fig1}
\end{center}
\end{figure}
In gravitational-wave interferometers composed of equal-length arms,  
the dynamics relevant to the output signal and the corresponding 
noise are described only by the antisymmetric mode of 
motion, $\widehat{x}$, of the four arm-cavity mirrors and by 
the dark-port sideband fields, 
which are decoupled from the other degrees of freedom~\cite{KLMTV}. 
In these devices, 
laser interferometry is used to monitor the displacement of the antisymmetric mode 
of the arm-cavity mirrors induced by the passage 
of a gravitational wave with (differential) strain $h$.
The output of the detector can be constructed from two
independent output observables, the two quadratures 
$\widehat{b}_1$ and $\widehat{b}_2$~\cite{KLMTV,BC2} (see Fig.~\ref{Fig1}) 
of the outgoing electromagnetic 
field immediately outside the SR mirror, which can be related 
to the input (noise) quadratures $\widehat{a}_1$, $\widehat{a}_2$ 
(see Fig.~\ref{Fig1}) and (the signal) $h$.  

Disregarding the motion of the mirrors during the 
light round-trip time (quasi-static approximation),
the radiation-pressure force acting on each arm-cavity mirror 
is $2 W/c$, where $W$ is the power circulating 
in each arm cavity, which is proportional to the square of the amplitude
of the electric field propagating toward the mirror and $c$ is the speed of 
light.

When the arm-cavity mirrors are held fixed, the
radiation-pressure force can be directly related to the dark-port 
quadrature fields~\cite{KLMTV}. 
In conventional interferometers such as LIGO-I, TAMA and VIRGO (see 
left panel on Fig.~\ref{Fig1}) the (Fourier domain) of this 
radiation-pressure force $\widehat{F}_0(\Omega)$ is determined
only by one of the input quadratures, say $\widehat{a}_1(\Omega)$~\cite{KLMTV}.
Since $[\widehat{a}_1(\Omega),\widehat{a}_1^\dagger(\Omega')] =0=
[\widehat{a}_2(\Omega),\widehat{a}_2^\dagger(\Omega')]$ and 
$[\widehat{a}_1(\Omega),\widehat{a}_2^\dagger(\Omega')]=2\pi i\,\delta(\Omega -\Omega')$,   
the response function of the {\it optical force} to 
perturbations caused by the mirror motion, 
which is given by $G_{FF}(t,t') \propto [\widehat{F}_0(t), 
\widehat{F}_0(t')]$, is zero. By contrast, in SR interferometers
such as LIGO-II (see right panel in Fig.~\ref{Fig1}), 
the radiation pressure force depends on a linear 
combination, with complex coefficients, of both the input 
quadratures $\widehat{a}_1(\Omega)$ 
and $\widehat{a}_2(\Omega)$. As a consequence, 
the response function $G_{FF}(t,t')\neq 0$. 
More specifically, taking into account the mirror motion, 
the full radiation-pressure force in SR interferometers, 
is given by~\cite{BC3}:   
\beq
\widehat{F}(t) = \widehat{F}_0(t) + \frac{i}{\hbar}
\,\int_{-\infty}^t d t'\,G_{FF}(t,t')\,\widehat{x}(t')\,.
\label{1}
\eeq
The second term in the RHS of the above equation can be 
easily explained in classical terms by noticing that the optical field 
fed back by the SR mirror into the arm cavities also contains
the classical GW signal $h$. Thus the radiation-pressure force 
$\widehat{F}$ must depend on the history of the 
antisymmetric mode of motion $\widehat{x}$.

\section{Dynamics, resonances and instability} 
\label{sec3}

In SR interferometers, the (Fourier domain) equation of motion 
for the antisymmetric mode of motion is~\cite{BC3}:
\beq
-\mu\,\Omega^2\,\widehat{x}(\Omega)= 
{\rm GW\,\, Force} + \widehat{F}_0(\Omega)
+ R_{FF}(\Omega)\,\widehat{x}(\Omega)\,,
\label{2}
\eeq
where $R_{FF}(\Omega)$ is the Fourier transform of the response function $G_{FF}$ 
and $\mu=m/4$ is the reduced mass of the antisymmetric mode, being  
$m$ the arm-cavity mirror mass.
Hence, from Eq.~(\ref{2}) we infer that the antisymmetric mode of motion 
is not only buffeted by the radiation pressure force $\widehat{F}_0$, 
but is also subject to a harmonic restoring force with 
frequency-dependent spring constant~\cite{BC3}:
\beq 
K(\Omega) = -R_{FF}(\Omega) \propto {I_o}\times 
({\rm SR\,mirror\, reflectivity}) \times({\rm SR\,detuning})\,,
\label{3}
\eeq
where $I_o$ is the laser light at the beamsplitter, and 
by SR detuning we mean the phase gained by the
laser carrier frequency in the SR cavity [see Refs.~\cite{BC2,BC3} for 
details]. This phenomenon, called ponderomotive rigidity, 
was originally discovered and analyzed in ``optical-bar'' 
GW detectors by Braginsky, Khalili and colleagues~\cite{OB}. 

In the absence of the SR mirror the optical-mechanical system 
formed by the optical fields and the arm-cavity mirrors is characterized 
by the mechanical (double) resonant frequency $\Omega_{\rm mech}^2=0$, 
related to the free motion of the antisymmetric mode, and by the 
optical resonant frequency $\Re{(\Omega_{\rm opt.})} =0$, $\Im{(\Omega_{\rm opt.})} =
-1/\tau_{\rm decay}$, where $\tau_{\rm decay}$ is the storage time of the arm cavity. 
When a highly reflecting SR mirror is added 
and we consider configurations with 
low light power, the optical field (almost) purely oscillates 
at the eingenfrequencies $\Omega_{\pm}$ at which the total round-trip 
phase in the entire cavity (arm cavity  + SR cavity) is $2\pi n$,  
with $n$ an integer. 

Since the ponderomotive rigidity $R_{FF} \propto {I_o}$, 
as we increase $I_o$ the test masses and the optical 
field get coupled more and more and we have a 
mixing of the mechanical and pure optical resonant frequencies. 
More specifically, the (coupled) mechanical resonance moves from zero 
as $\sim I_o^{1/2}$, while the (coupled) optical resonances get shifted 
away from the values $\Omega_\pm$ as $\sim I_o$. 

We have found~\cite{BC3} that the (coupled) mechanical 
resonant frequencies have always a positive imaginary part, 
corresponding to an instability. 
This instability has an origin similar to the dynamical
instability induced in a detuned Fabry-Perot cavity by
the radiation-pressure force acting on the mirrors~\cite{All,OB}.
To suppress this instability, we proposed a
feed-back control system that does not compromise the GW interferometer 
sensitivity. However, although the model we used to describe the servo system~\cite{BC3} 
may be realistic for an all-optical control loop, this might not be the case if an electronic 
servo system is implemented. Thus, a more 
thorough formulation should be used to fully describe
this latter case~\cite{BCM}.

\section{Quantum-noise spectral density}
\label{sec4}

In light of the discussion at the end of last section,  
let us derive the noise spectral density of a (stabilized) 
interferometer~\cite{BC3}. To identify the radiation pressure and the shot noise 
contributions in the total optical noise, we use the fact that they transform  
differently under rescaling of the reduced mass $\mu$. 
Indeed, it is straightforward to show~\cite{BC2} that in the 
total optical noise there exist only two kinds of terms. 
There are terms that are invariant under rescaling of $\mu$ and terms 
that are proportional to $1/\mu$. Quite generally 
we can rewrite the (Fourier domain) output $\widehat{\cal O}$ as~\cite{BC1}:
\beq
\widehat{\cal O}(\Omega) = \widehat{\cal Z}(\Omega) + {\cal R}_{xx}(\Omega)\,
\widehat{\cal F}(\Omega) + L\,h(\Omega)\,,
\label{4}
\eeq
where by output we mean (modulo a normalization factor) 
one of the two (stabilized) quadratures $\widehat{b}_1$, $\widehat{b}_2$~\cite{BC3} 
or a combination of them. In Eq.~(\ref{4}) ${\cal R}_{xx}=
-{1}/{\mu\,\Omega^2}$ is the susceptibility of the antisymmetric mode 
of motion of the four arm-cavity mirrors and $L$ is the arm-cavity length. 
The observables $\widehat{\cal Z}$ and $\widehat{\cal F}$ 
do not depend on the mirror masses $\mu$~\cite{BC2}, and 
we refer to them as the {\em effective} shot 
noise and {\em effective} radiation-pressure force, 
respectively. The (one-sided) noise spectral density reads~\cite{BK92}: 
\beq
S_{h}(\Omega)=\frac{1}{L^2}\,\{S_{\widehat{\cal Z} \widehat{\cal Z}}(\Omega)
      +2{\cal R}_{xx}(\Omega)\,\Re[S_{\widehat{\cal F} \widehat{\cal Z}}(\Omega)] 
+ {\cal R}_{xx}^2(\Omega)\, S_{\widehat{\cal F} \widehat{\cal F}}(\Omega)\}\,,
\label{5}
\eeq
where we defined $
2\pi\,\delta\left(\Omega-\Omega'\right)\,S_{\widehat{\cal A} \widehat{\cal B}}(\Omega)= 
\langle \widehat{\cal A}(\Omega) \widehat{\cal B}^\dagger(\Omega')+ \widehat{\cal B}^\dagger
(\Omega') \widehat{\cal A}(\Omega)\rangle$. Moreover, the 
(one-sided) spectral densities and cross correlations of $\widehat{\cal Z}$ and 
$\widehat{\cal F}$ satisfy the uncertainty relation~\cite{BK92}:
\beq
S_{\widehat{\cal Z} \widehat{\cal Z}}(\Omega)\,S_{\widehat{\cal F} 
\widehat{\cal F}}(\Omega) - 
S_{\widehat{\cal Z} \widehat{\cal F}}(\Omega)\,S_{\widehat{\cal F} \widehat{\cal Z}}(\Omega) \ge \hbar^2\,.
\label{6}
\eeq
It is possible to show~\cite{BC3} that the ponderomotive effect discussed 
in Sec.~\ref{sec3}, can be directly related to the presence of 
dynamical correlations between the shot-noise and radiation-pressure 
noise~\cite{FK,BC4}. 

In conventional interferometers such as 
LIGO-I, TAMA and VIRGO, the ponderomotive effect is absent, i.e.\ $R_{FF}=0$. 
In this case, as long as squeezed-input light is not 
injected into the interferometer from the dark-port and/or correlations are not built 
up statically during the readout process~\cite{SHFD,KLMTV}, we have 
$S_{\widehat{\cal Z} \widehat{\cal F}}=0=S_{\widehat{\cal F}
\widehat{\cal Z}}$. Thus, in conventional interferometers 
Eq.~(\ref{6}) imposes the following lower bound on 
the noise spectral density: $S_{h}^{\rm conv.}(\Omega)
\ge S_h^{\rm SQL}(\Omega)\equiv {2 \hbar}/{\mu\,\Omega^2\,L^2}$. 
The quantity $S_h^{\rm SQL}(\Omega)$ is generally called 
the standard quantum limit (SQL) for the dimensionless 
GW signal $h =\Delta L/L$.

\begin{figure}
\begin{center}
\epsfig{file=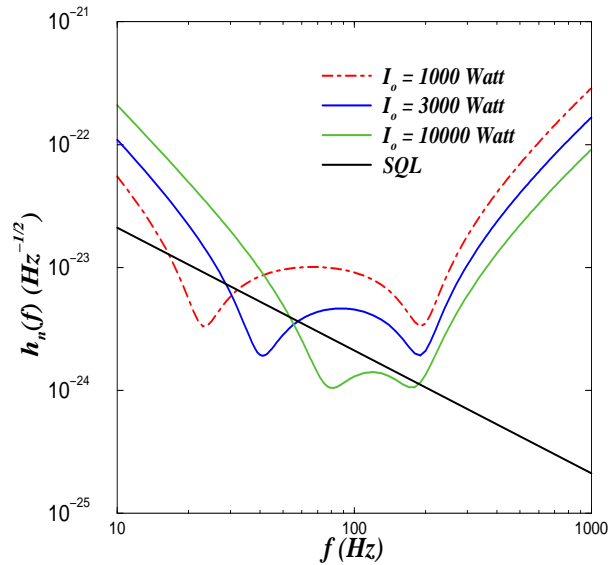,width=0.5\textwidth,height=0.5\textwidth,angle=-90} 
\caption{Plot of the square root of the quantum-noise spectral density $h_n \equiv 
\sqrt{S_h}$ versus frequency, for various choices of the light-power 
at the beamsplitter, having fixed the SR mirror reflectivity 
and the SR detuning. The SQL line is also shown.}
\label{Fig2}
\end{center}
\end{figure}

In SR interferometers, and ``optical-bar'' GW detectors as well~\cite{OB}, 
because of the ponderomotive effect 
($R_{FF}\neq 0$) shot-noise and radiation-pressure noise 
are automatically correlated and Eq.~(\ref{6}) does no longer impose a 
lower bound on the noise spectral density Eq.~(\ref{5}). 
In particular, we found~\cite{BC2} that there exists an experimentally
accessible region of the parameter space for which 
the quantum noise curves can beat the SQL by roughly 
a factor of two over a bandwidth $\Delta f \! \sim \! f$.
This fact is illustrated in Fig.~\ref{Fig2}, where the 
square root of the noise spectral density ($h_n \equiv 
\sqrt{S_h}$) is plotted 
versus frequency, for various choices of the light power 
at the beamsplitter, having fixed the SR mirror reflectivity 
and the SR detuning. Note the two distinct valleys which 
go below the SQL line. Their position is determined by the (coupled) 
resonant frequencies of the optical-mechanical 
system discussed in Sec.~\ref{sec3}. As anticipated in 
the previous section, as we increase $I_o$ the (coupled) 
mechanical resonant frequency (on the left) moves from zero to 
the right, while the (coupled) optical resonant frequency (on the right) does 
not vary much, being present already as pure optical resonance 
in the limit of low light power.

The total noise, which includes seismic, suspension and 
thermal contributions, can beat the SQL only if all other noise 
sources can also be pushed below the SQL. 
These noises are not quantum limited in principle but may be technically challenging to
reduce~\cite{GSSW99,BGV00}.

\section{Conclusions}
\label{sec5}

Our analyses~\cite{BC1,BC2,BC3} have revealed that in SR interferometers, 
the dynamics of the whole optical-mechanical system, composed of 
the arm-cavity mirrors and the optical field, resembles 
that of a free test mass (mirror motion) connected to a massive spring (optical fields). 
When the test mass and the spring are not connected (e.g., for very low laser power) 
they have their own eigenmodes, namely the uniform
translation mode for the free antisymmetric mode, and the 
longitudinal-wave mode for the spring 
(decoupled SR optical resonance). However, for LIGO-II laser power 
the test mass is connected to the massive spring and 
the two free modes become shifted in frequency, so the entire coupled
system can resonate at two pairs of finite frequencies.
Near these resonances the noise curve can 
beat the free mass SQL, as shown in Fig.~\ref{Fig2}.
This phenomenon is not unique to SR interferometers; it 
is a generic feature of detuned cavities \cite{All,FK,BC4} and 
was used by Braginsky, Khalili and colleagues in 
designing the ``optical bar'' GW detectors~\cite{OB}.

However, the optical-mechanical system is by itself dynamically unstable, 
and a much more careful and precise study of the control system 
should be carried out, including various readout schemes~\cite{BCM},
before any practical implementation.

\vspace{-0.5cm}

\ack
\vspace{-0.3cm}
This research was supported by NSF
grants PHY-9900776 and PHY-0099568 
and for AB also by Caltech's Richard C Tolman Fellowship.

\end{document}